\documentclass{ieeeaccess}
\usepackage{cite}
\usepackage{amsmath,amssymb,amsfonts}
\usepackage{algorithmic}
\usepackage{graphicx}
\usepackage{textcomp}
\usepackage[pagewise,left]{lineno}

\usepackage{float}
\usepackage{array}
\usepackage{multirow}
\newcolumntype{C}[1]{>{\centering\arraybackslash}m{#1}}

\ifCLASSINFOpdf
\def\BibTeX{{\rm B\kern-.05em{\sc i\kern-.025em b}\kern-.08em
    T\kern-.1667em\lower.7ex\hbox{E}\kern-.125emX}}
\begin{document}
\history{}
\doi{}

\title{A two-stage approach for beam hardening artifact reduction in low-dose dental CBCT}
\author{\uppercase{Taigyntuya Bayaraa\authorrefmark{1}, Chang Min Hyun\authorrefmark{1}, Tae Jun Jang\authorrefmark{1}, Sung Min Lee\authorrefmark{2}, and Jin Keun Seo\authorrefmark{1}, \IEEEmembership{Member, IEEE}}}
\address[1]{School of Mathematics and Computing (Computational Science and Engineering), Yonsei University, Seoul, 03722, South Korea}
\address[2]{Software Division, HDXWILL, Seoul, 08501, South Korea}

\tfootnote{This work was supported by Samsung Science \& Technology Foundation (No. SRFC-IT1902-09).}

\corresp{Corresponding author: Chang Min Hyun (email: chammyhyun@yonsei.ac.kr).}

\begin{abstract}
This paper presents a two-stage method for beam hardening artifact correction of dental cone beam computerized tomography (CBCT).
The proposed artifact reduction method is designed to improve the quality of maxillofacial imaging, where soft tissue details are not required.
Compared to standard CT, the additional difficulty of dental CBCT comes from the problems caused by offset detector, FOV truncation, and low signal-to-noise ratio due to low X-ray irradiation.
To address these problems, the proposed method primarily performs a sinogram adjustment in the direction of enhancing data consistency, considering the situation according to the FOV truncation and offset detector.
This sinogram correction algorithm significantly reduces beam hardening artifacts caused by high-density materials such as teeth, bones, and metal implants, while tending to amplify special types of noise.
To suppress such noise, a deep convolutional neural network is complementarily used, where CT images adjusted by the sinogram correction are used as the input of the neural network.
Numerous experiments validate that the proposed method successfully reduces beam hardening artifacts and, in particular, has the advantage of improving the image quality of teeth, associated with maxillofacial CBCT imaging.
\end{abstract}


\titlepgskip=-15pt

\maketitle

\section{Introduction}
In clinical dentistry, dental cone beam computerized tomography(CBCT) has been gaining significant attention as a crucial supplement radiographic technique to aid diagnosis, treatment planning, and prognosis assessment such as diagnosis of dental caries, reconstructive craniofacial surgery planning, and evaluation of the patient's face \cite{gupta2013,miracle2009,sukovic2003,scarfe2017}.
In particular, with relatively low-dose radiation exposure, dental CBCT allows the provision of high quality 3D maxillofacial images, which can be used in a wide range of clinical applications in order to understand the complicated anatomical relationships and the surrounding information of the maxillofacial skeleton.
Nevertheless, maxillofacial CBCT imaging still suffers from various artifacts that significantly degrade the image quality regarding bone and teeth. Compared to standard multi-detector CT (MDCT), the additional difficulty of artifact reduction in most dental CBCT is caused by the use of low X-ray irradiation and a small-size flat-panel detector in which the center axis of rotation is offset relative to the source-detector axis to maximize the transaxial FOV \cite{chang1995,cho1995}.

As the number of patients with metallic implants and dental filling is increasing, metal-induced artifacts are common in dental CBCT \cite{draenert2007,razavi2010,sanders2007,schulze2010}.
These metal-related artifacts are generated by the effects of beam hardening-induced sinogram inconsistency and different types of complicated metal-bone-tissue interactions with factors such as scattering, nonlinear partial volume effects, and electric noise \cite{hsieh2002,barrett2004,schulze2011,lee2019}. Furthermore, reducing metal-induced artifacts, which is known to be a very challenging problem in all kinds of CT imaging \cite{de1998,gjesteby2016}, is much difficult in the dental CBCT environment owing to additional problems arising from offset detectors, FOV truncations, and low X-ray doses.

\begin{figure*}[t]
	\centering
	\includegraphics[width=0.9\textwidth]{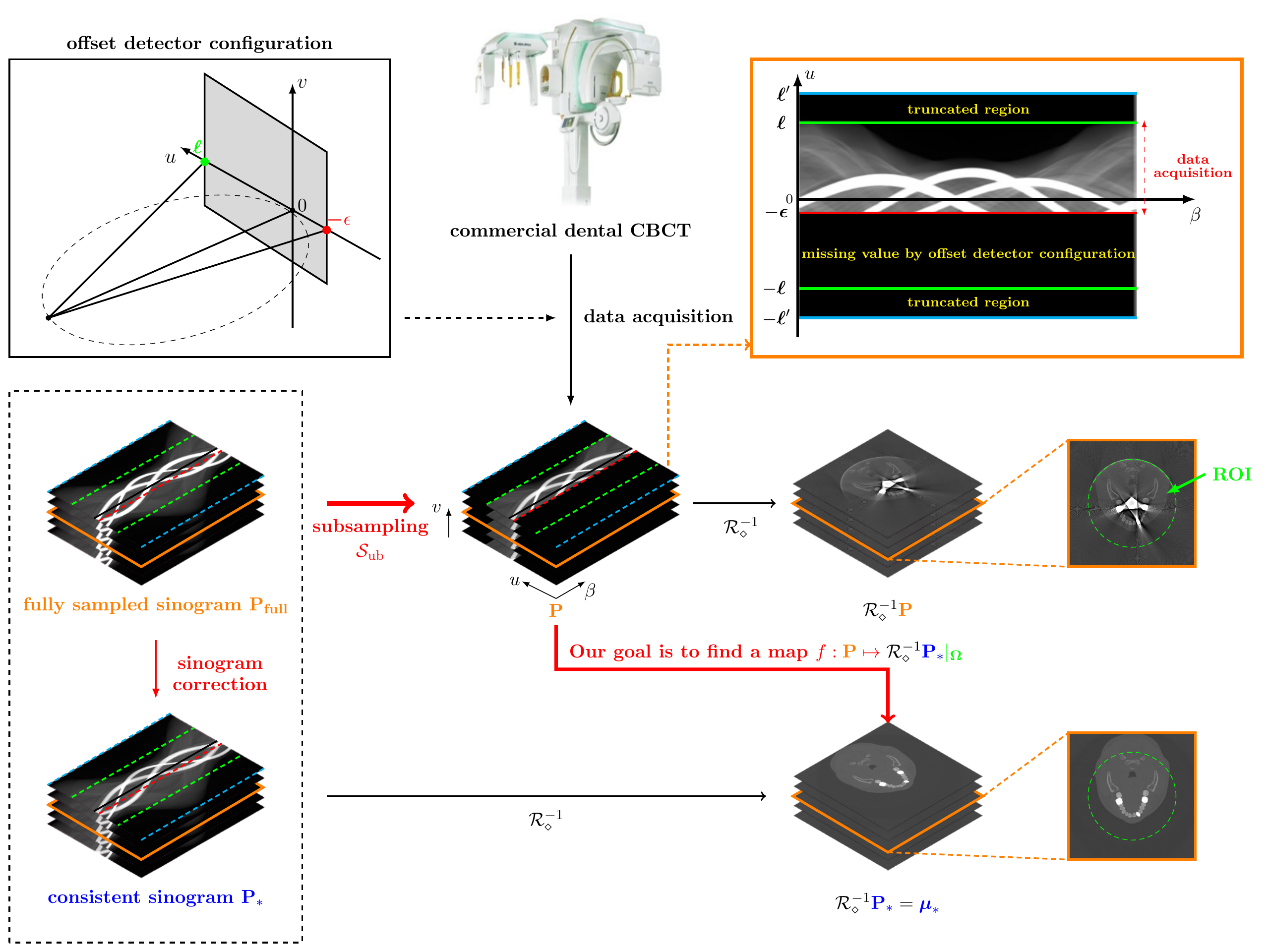}
	\caption{Beam hardening artifact reduction in dental CBCT is to find a reconstruction function $f$ that recovers a local ROI $\boldsymbol \Omega$ of a desired beam hardening artifact-free image $\mathcal R_{\diamond}^{-1} \mbox{\textbf{P}}_*$ from a dental CBCT sinogram $\mbox{\textbf{P}}$. Owing to the offset detector configuration and interior-ROI-oriented scan, $\mbox{\textbf{P}}$ can be viewed as a subsampled sinogram from a fully sampled sinogram $\mbox{\textbf{P}}_{\mbox{\tiny full}}$ acquirable in the standard CBCT. This subsampling causes sinogram truncation and asymmetry.}
	\label{offsetCBCTsystem}
\end{figure*}

There have been extensive research efforts for beam-hardening artifact correction (BHC), which reduce metal-induced streaking and shadow artifacts without affecting intact anatomical image information.
In the dental CBCT environment, the existing BHC methods are not applicable, do not reduce the metal artifacts effectively, may introduce new streaking artifacts that didn't previously exist, or require huge computational complexity.
Dual-energy CT \cite{alvarez1976,lehmann1981,yu2012} requires a higher dose of radiation compared with single-energy CT \cite{casteele2004}; therefore, this approach is not suitable for low-dose dental CBCT. In raw data correction methods, unreliable background data due to the presence of metallic objects can be recovered using various inpainting techniques such as interpolation \cite{abdoli2010,bazalova2007,kalender1987,lewitt1978,roeske2003}, normalized interpolation (NMAR) \cite{Meyer2010}, Poisson inpainting \cite{Park2013},  wavelet \cite{Mehranian2013,zhao2000,zhao2001}, tissue-class model \cite{Bal2006}, total variation \cite{duan2008}, and Euler's elastica \cite{gu2006}.
These methods might introduce new artifacts that did not previously exist.
Moreover, these techniques tend to impair the morphological information in the areas around the metal objects in the reconstructed images.
Various iterative reconstruction methods have been developed for BHC \cite{de2001,elbakri2002,menvielle2006,sullivan2007,williamson2002}.
This approach requires extensive knowledge about the CT system configuration, and the associated computation time for full iterative reconstructions can be clinically prohibitive.

\begin{figure*}[t!]
	\centering
	\includegraphics[width=1\textwidth]{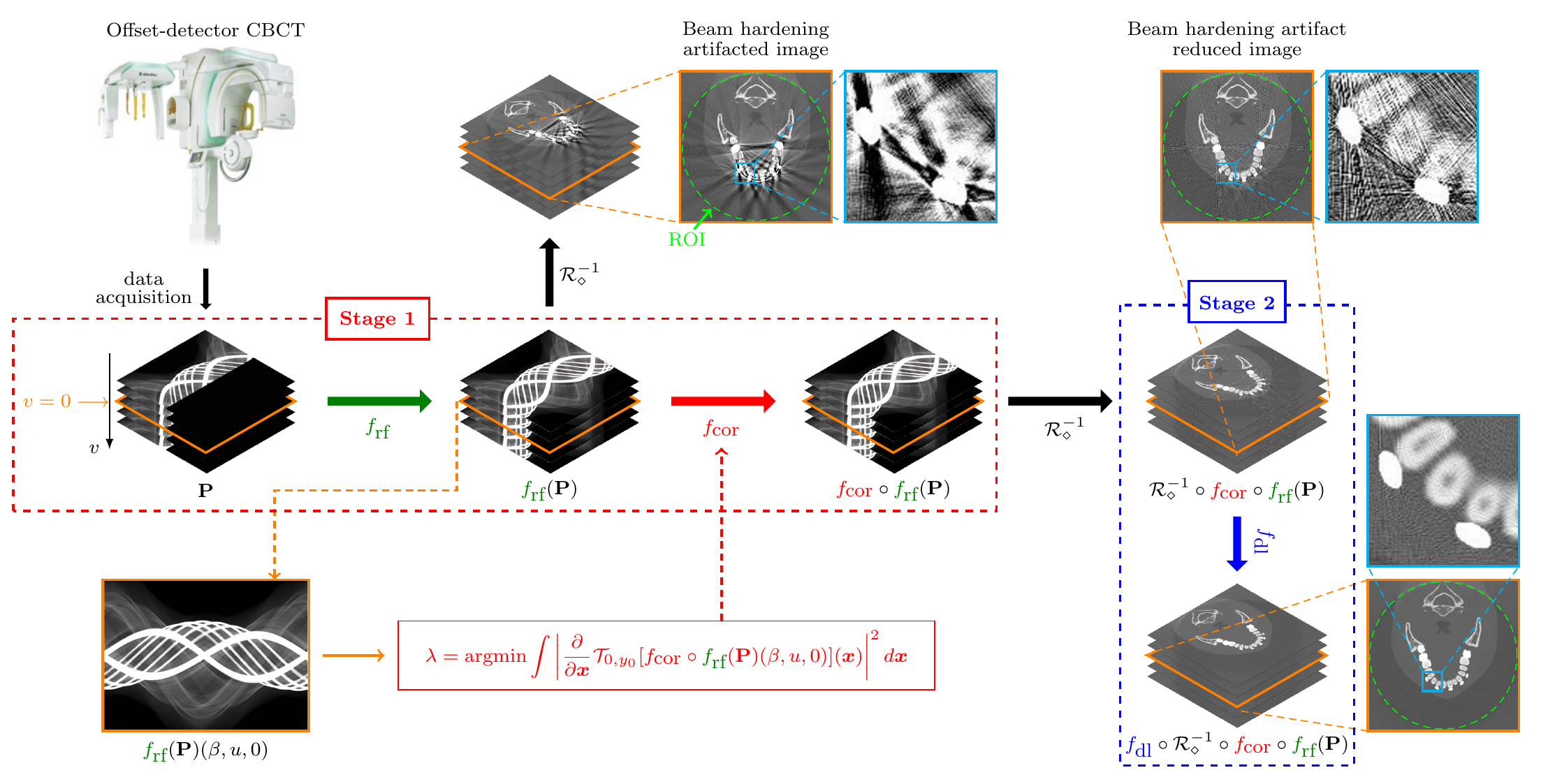}
	\caption{The proposed two stage method for beam hardening artifact reduction in dental CBCT. The proposed method comprises (stage 1) sinogram reflection and sinogram adjustment and (stage 2) deep learning. The sinogram reflection $f_{\mbox{\tiny rf}}$ is used to recover a missing data in $\mbox{\textbf{P}}$ caused by the offset detector acquisition. After then, the proposed method alleviates the beam hardening-induced sinogram inconsistency by applying the sinogram adjustment algorithm $f_{\mbox{\tiny cor}}$. Using a trained deep learning network $f_{\mbox{\tiny dl}}$. the reconstruction image is further improved.}
	\label{overall_sinocor}
\end{figure*}

A direct sinogram inconsistency correction method \cite{lee2019} was proposed recently to alleviate beam-hardening factors, while keeping a part of the data where beam hardening effects are small. This approach has an advantage over conventional image processing-based methods in that it does not require any segmentation of the metal region. Unfortunately, this method cannot be directly applied in our case because of problems caused by offset detector and FOV truncation. In \cite{park2018}, a deep learning-based sinogram correction method is used to reduce the primary metal-induced beam-hardening factors along the metal trace in the sinogram. This method was applied to the restricted situation of patient-implant-specific model, where the mathematical beam hardening corrector \cite{park2016,park2017} of a given metal geometry effectively generates simulated training data. This type of approaches may not be suitable for the dental CBCT case, in which the sinogram mismatch is intertwined by complex factors associated with various geometry of metals, metal-bone and metal-teeth interaction, FOV truncation, offset detector acquisition, and so on.

To overcome the aforementioned difficulties arising in low-dose offset-detector CBCT, this paper proposes a two-stage BHC method that mainly focuses on improving the quality of maxillofacial imaging, where soft tissue details are not required. In the first stage, we apply sinogram inconsistency correction by adjusting the sinogram intensity to reveal the anatomical structure obscured by the artifact. To perform the sinogram adjustment under the offset-detector CBCT environment, the proposed method uses a sinogram reflection technique and the data consistency condition \cite{clackdoyle2015} related to the sinogram consistency in the presence of FOV truncation. In the second stage, a supplemental deep learning technique is employed to eliminate the remaining streaking artifacts.

Numerical simulations and real phantom experiments show that the proposed method effectively reduces beam hardening artifacts and offers the benefit of improving the image quality of bone and teeth associated with maxillofacial CBCT imaging.

\begin{figure*}[t!]
	\centering
	\includegraphics[width=0.95\textwidth]{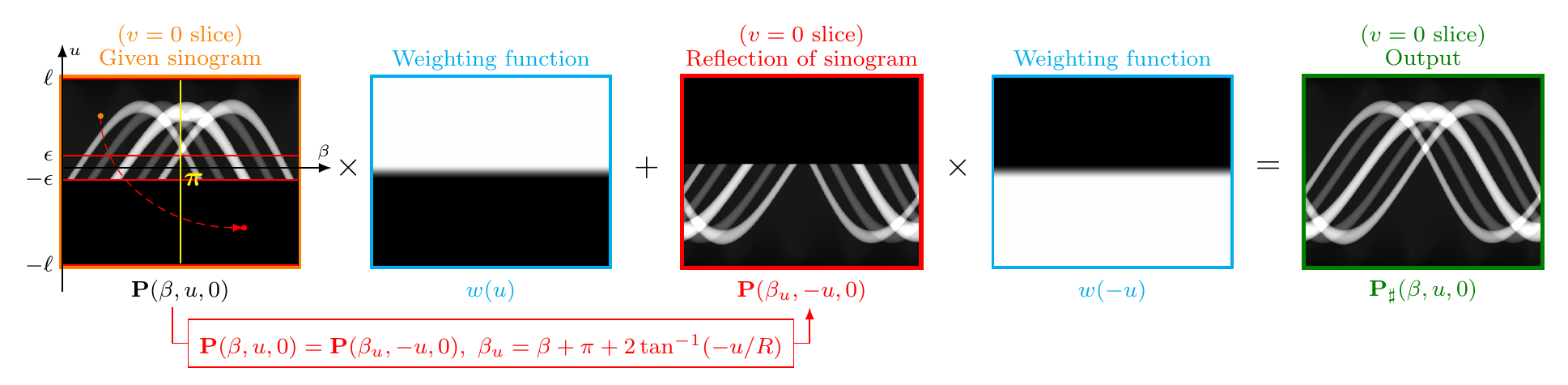}
	\caption{The sinogram reflection process. The missing part of \mbox{\textbf{P}} is recovered using the relation \eqref{reflec} and a sinogram $\mbox{\textbf{P}}$. The filled sinogram $\mbox{\textbf{P}}_{\sharp}$ is obtained based on the relation (\ref{weighting}). This process is repeated in all sinogram slices.}
	\label{refalg}
\end{figure*}
\begin{figure*}[t]
	\centering
	\includegraphics[width=0.9\textwidth]{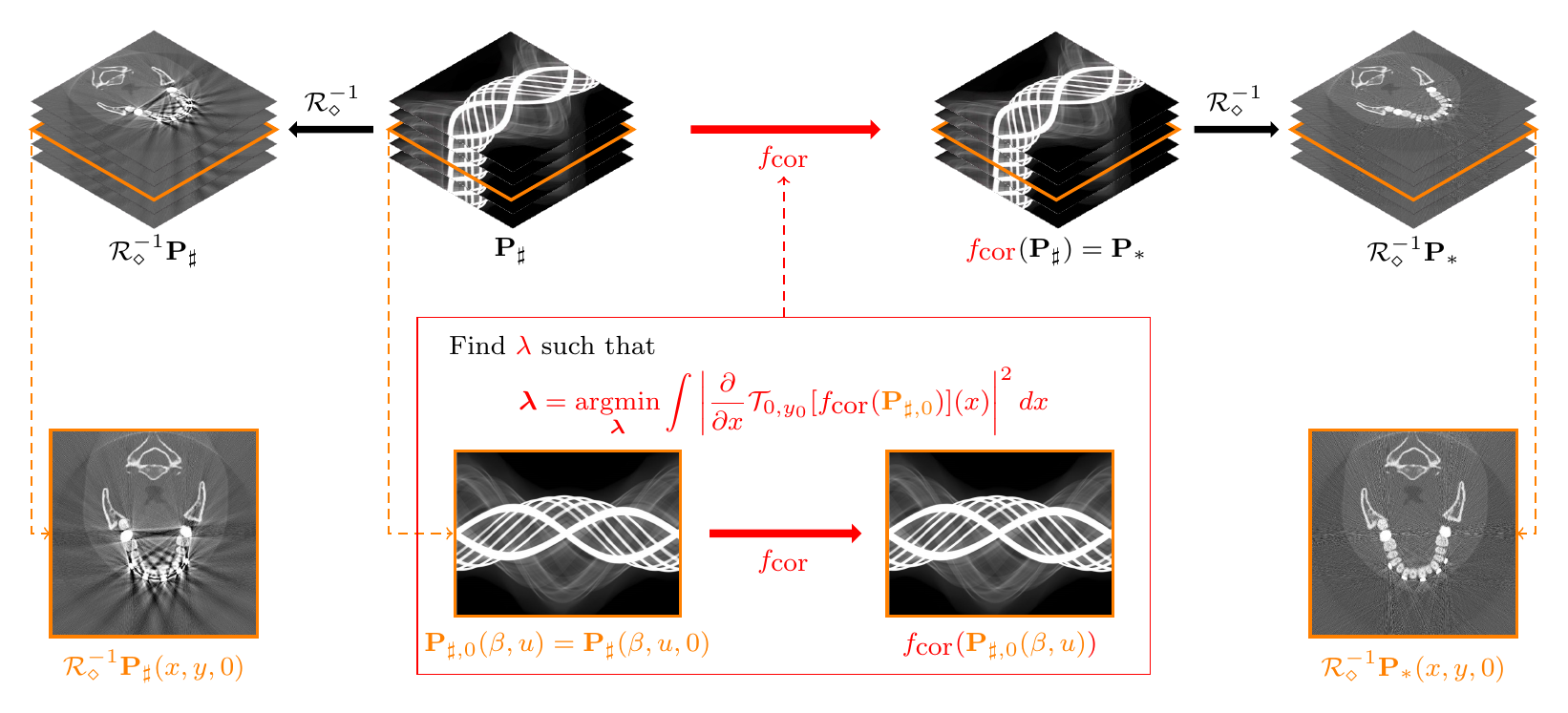}
	\caption{Sinogram inconsistency correction process. To determine an unknown set of parameters $\boldsymbol \lambda$ in the correction function $f_{\mbox{\tiny cor}}$, the optimization procedure involving the data consistency condition taking the situation of FOV truncation into consideration is used with a sinogram slice $\mbox{\textbf{P}}_\sharp(\beta,u,0)$. After finding $\boldsymbol \lambda$, a whole sinogram is adjusted by $f_{\mbox{\tiny cor}}$.}
	\label{sino_incon_corr}
\end{figure*}

\section{Method}
The proposed method is for the most widely used dental CBCT systems shown in Figure \ref{offsetCBCTsystem}, which use an offset detector configuration and an interior-ROI-oriented scan. Let $\mbox{\textbf{P}}(\beta,u,v)$ denote an acquired sinogram, where $\beta \in [0,2\pi)$ is the projection angle of the X-ray source rotated along the circular trajectory, and $(u,v)$ is the coordinate system of the 2D flat-panel detector. Because the effective FOV does not cover the entire region of an object to be scanned, the sinogram $\mbox{\textbf{P}}$ can be expressed by
\begin{equation} \label{pfull}
\mbox{\textbf{P}} = \mathcal S_{\mbox{\scriptsize ub}}( \mbox{\textbf{P}}_{\mbox{\scriptsize full}})
\end{equation}
where $\mbox{\textbf{P}}_{\mbox{\scriptsize full}}$ is the corresponding sinogram acquirable with non-offset and wide-detector CBCT providing a whole information of a sinogram and $\mathcal S_{\mbox{\scriptsize ub}}$ is a subsampling operator determined by the size and offset configuration of a detector. More precisely, let a 2D flat-panel detector be aligned in $[-\epsilon, \ell]$ with respect to $u$-axis. As shown in Figure \ref{offsetCBCTsystem}, sinogram $\mbox{\textbf{P}}$ is truncated by
\begin{equation}
\mbox{\textbf{P}} = \mathcal S_{\mbox{\scriptsize ub}}( \mbox{\textbf{P}}_{\mbox{\scriptsize full}}) =  \left\{ \begin{array}{cl} \mbox{\textbf{P}}_{\mbox{\scriptsize full}} & \mbox{ if } u \in [-\epsilon, \ell] \\ 0 & \mbox{ if } u \in [-\ell^\prime, -\epsilon] \cup [\ell, \ell^\prime]  \end{array} \right.
\end{equation}
where $[-\ell^\prime, \ell^\prime]$ is the support of $\mbox{\textbf{P}}_{\mbox{\scriptsize full}}$ with respect to the $u$-axis.
This missing information in $\mbox{\textbf{P}}$ along the $u$-axis makes the application of existing methods difficult.

Metal-related beam hardening artifacts are caused by the polychromatic nature of the X-ray source beam. According to the Beer-Lambert law \cite{beer1852}, the sinogram $\mbox{\textbf{P}}$ is given by
\begin{equation}
\mbox{\textbf{P}} = \mathcal S_{\mbox{\scriptsize ub}}( -\mbox{ln}\int_{E} \eta(E) \exp(-\mathcal R_{\diamond} \boldsymbol \mu_{E})dE )
\end{equation}
where $\mathcal R_{\diamond}$ is a cone beam projection associated with 3D Radon transform, and $\boldsymbol \mu_{E}$ is a three-dimensional distribution at an energy level $E$.
In the presence of high-attenuation objects such as metal, the sinogram inconsistency between $\mbox{\textbf{P}}$ and the reconstruction model (based on the assumption of the monochromatic X-ray beam) generates streaking and shadowing artifacts in a reconstructed image \cite{park2017}.

As described in Figure \ref{overall_sinocor}, the proposed method is composed of the following four functions:
\begin{equation} \label{recon}
f = f_{\mbox{\scriptsize dl}} \circ \mathcal R_{\diamond}^{-1} \circ f_{\mbox{\scriptsize cor}} \circ f_{\mbox{\scriptsize rf}}
\end{equation}
where
\begin{itemize}
	\item $f_{\mbox{\scriptsize rf}}$ is a sinogram reflection process, which estimates missing data in $\mbox{\textbf{P}}$ by the offset detector configuration (see Figure \ref{refalg}).
	\item $f_{\mbox{\scriptsize cor}}$ is a sinogram inconsistency corrector, which alleviates a beam hardening-induced sinogram inconsistency while considering FOV truncation (see Figure \ref{sino_incon_corr}).
	\item $\mathcal R_{\diamond}^{-1}$ is a standard FDK algorithm \cite{feldkamp1984} with the addition of the sinogram extrapolation method \cite{tisson2006}.
	\item $f_{\mbox{\scriptsize dl}}$ is a deep learning network, which further improves the reconstruction image (see Figure \ref{dl_process}).
\end{itemize}
The proposed method is designed to generate a reconstruction function $f : \mbox{\textbf{P}} \mapsto  \mathcal R^{-1}_{\diamond} \mbox{\textbf{P}}_* |_{\mbox{\scriptsize ROI}}$, where $\mathcal R^{-1}_{\diamond} \mbox{\textbf{P}}_* |_{\mbox{\scriptsize ROI}}$ represents a local ROI reconstruction with $\mbox{\textbf{P}}_*$  being a corrected sinogram of $\mbox{\textbf{P}}$. Here, ROI is the region of interest that is determined by the truncated sinogram. The ROI is represented at the right-middle side of Figure \ref{offsetCBCTsystem}. The sinogram $\mbox{\textbf{P}}_{*}$ can be $\mbox{\textbf{P}}_{*}=\mathcal R_{\diamond} \boldsymbol \mu_*$ for an attenuation coefficient distribution $\boldsymbol \mu_*$ at a mean energy level $E_*$. The following subsections explain each process in detail.

\begin{figure*}[t]
	\centering
	\includegraphics[width=0.9\textwidth]{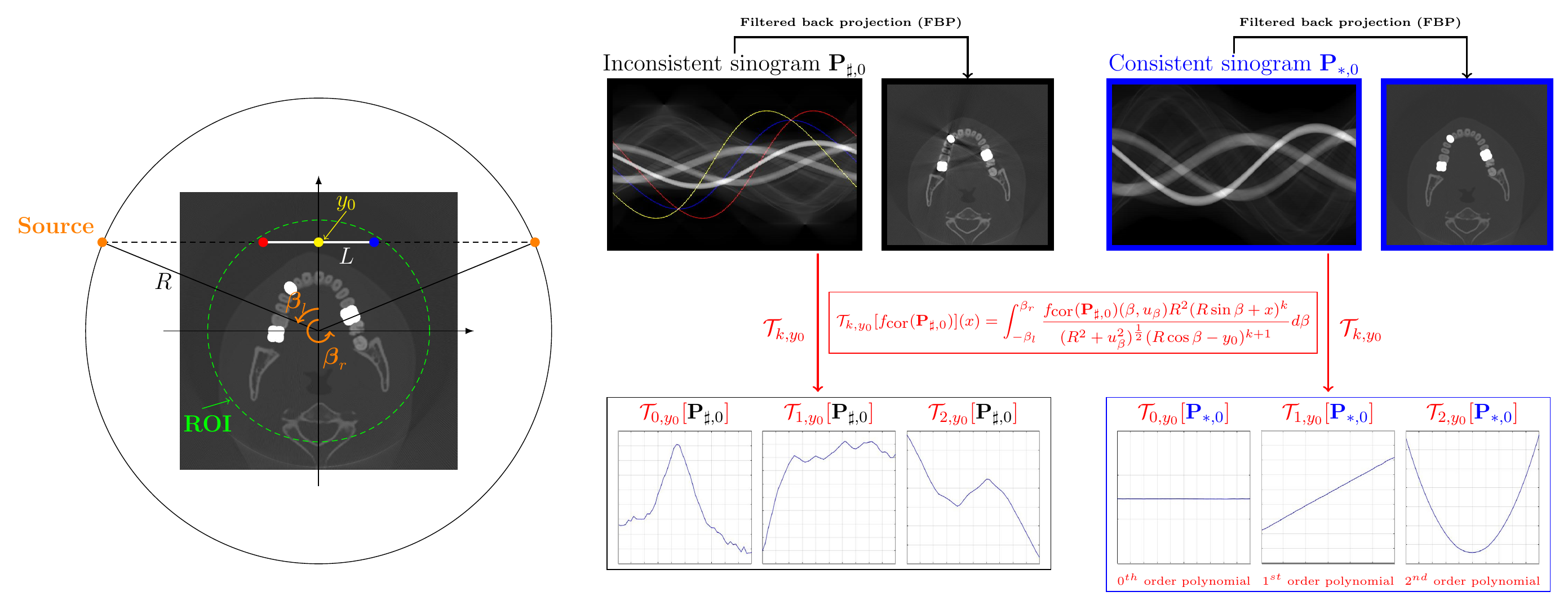}
	\caption{Data consistency condition for a truncated fan-beam sinogram data. For a fan beam sinogram $\mbox{\textbf{P}}_{\sharp,0}$ possessing truncation issue and any nonzero integer $k$, if $\mbox{\textbf{P}}_{\sharp,0}$ is consistent, the function $\mathcal T_{k,y_0}[\mbox{\textbf{P}}_{\sharp,0}]$ becomes a polynomial with degree $k$.}
	\label{cons_trun}
\end{figure*}
\subsubsection{Sinogram reflection}
The sinogram reflection process provides the missing part of $\mbox{\textbf{P}}$ (i.e. missing information in $u \in [-\ell, -\epsilon]$). This filling should be based on the following approximate identity of $\mbox{\textbf{P}}_{\mbox{\scriptsize full}}$ in \eqref{pfull}:
\begin{equation} \label{reflec}
\mbox{\textbf{P}}_{\mbox{\scriptsize full}}(\beta,u,v) \approx \mbox{\textbf{P}}_{\mbox{\scriptsize full}}(\beta_{u}, -u, v), ~ \forall ~ u \in [0,\ell]
\end{equation}
where $R$ is the distance from th X-ray source to the isocenter and
\begin{equation}
\beta_{u}=\beta + \pi + 2 \tan^{-1}(-\dfrac{u}{R})
\end{equation}
Note that the above approximation becomes the equality as $v \rightarrow 0$.

Based on \eqref{reflec}, the filled sinogram $\mbox{\textbf{P}}_{\sharp}$ is obtained as follows:
\begin{align}\label{weighting}
\mbox{\textbf{P}}_\sharp(\beta,u,v) = \left\{ \begin{array}{ll} \mbox{\textbf{P}}(\beta, u , v) &  \mbox{ if } u \in [\epsilon, \ell] \vspace{0.2cm} \\
 \begin{array}{l} \hspace{-0.2cm} \omega(u)\mbox{\textbf{P}}(\beta_{u},-u,v) \\ \hspace{-0.2cm}  +(1-\omega(u))\mbox{\textbf{P}}(\beta,u,v)  \end{array} \vspace{0.2cm} &  \mbox{ if } u \in (-\epsilon, \epsilon)  \\ \mbox{\textbf{P}}(\beta_{u},-u , v) &   \mbox{ if } u \in [-\ell, -\epsilon]  \end{array} \right.
\end{align}
where $\omega$ is a weighting function given by
\begin{equation}
\omega(u) = \dfrac{1-\cos(\pi(-u+\epsilon)/(2\epsilon))}{2}
\end{equation}
The function $f_{\mbox{\scriptsize rf}}$  in \eqref{recon} is the map from $\mbox{\textbf{P}}$ to $\mbox{\textbf{P}}_{\sharp}$.

\subsubsection{Sinogram inconsistency correction} \label{secSIC}
The sinogram inconsistency correction alleviates the beam hardening-induced sinogram inconsistency in $\mbox{\textbf{P}}_\sharp$ by developing the sinogram inconsistency corrector $f_{\mbox{\scriptsize cor}}$. The goal is to find the corrector function $f_{\mbox{\scriptsize cor}} : \mbox{\textbf{P}}_\sharp \mapsto \mbox{\textbf{P}}_*$ that maps from the inconsistent sinogram $ \mbox{\textbf{P}}_\sharp$ to a consistent sinogram $\mbox{\textbf{P}}_*$, which lies in the range space of the CBCT model. Note that the corrector acts in the restricted interval $u \in [-\ell, \ell]$.

\begin{figure*}[t]
	\centering
	\includegraphics[width=0.9\textwidth]{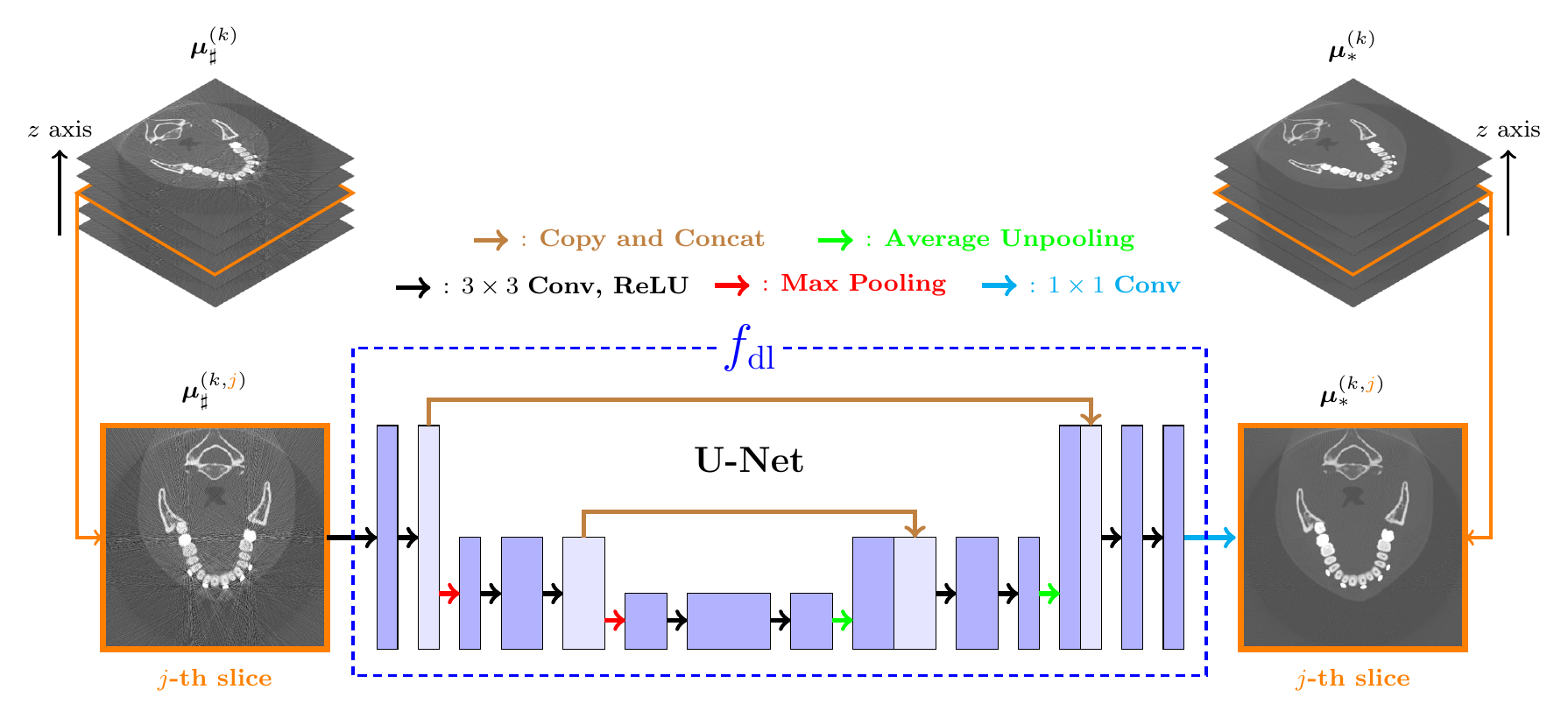}
	\caption{Deep Learning process is used for improving the quality of a reconstructed image. Using a training data set $\{(\boldsymbol \mu_\sharp^{(k,j)},\boldsymbol \mu_{*}^{(k,j)})\}_{k,j}$, the convolutional neural network U-net learns a function $f_{\mbox{\tiny dl}}$ that maps a reconstructed image $\boldsymbol \mu_\sharp$ by sinogram adjustment algorithm to our desired image $\boldsymbol \mu_{*}$.}
	\label{dl_process}
\end{figure*}

The proposed method is based on the following polynomial approximation:
\begin{equation} \label{poly}
	\mbox{\textbf{P}}_* \approx f_{\mbox{\scriptsize cor}}(\mbox{\textbf{P}}_\sharp) =  \sum_{i=0}^{n} \lambda_i \mbox{\textbf{P}}_\sharp^i
\end{equation}
where the coefficients $\lambda_0,\cdots,\lambda_n$ are determined in the way that $ f_{\mbox{\scriptsize cor}}(\mbox{\textbf{P}}_{\sharp})$ satisfies the data consistency condition \cite{clackdoyle2015}. To be precise, $\boldsymbol \lambda = (\lambda_0,\cdots,\lambda_n)$ can be determined by solving the following minimization problem:
\begin{equation} \label{mini_para}
\boldsymbol \lambda = \underset{\boldsymbol \lambda}{\mbox{argmin}} \sum_{k=0}^{k_*} \int
\left| \dfrac{\partial^{k+1}}{\partial x^{k+1}} \mathcal T_{k,y_0}[f_{\mbox{\scriptsize cor}}(\mbox{\textbf{P}}_{\sharp,0})](x) \right|^2 dx
\end{equation}
where $\mbox{\textbf{P}}_{\sharp,0}(\beta,u)=\mbox{\textbf{P}}_{\sharp}(\beta,u,0)$ and
\begin{equation} \label{consistency}
\small{\mathcal T_{k,y_0}[f_{\mbox{\scriptsize cor}}(\mbox{\textbf{P}}_{\sharp,0})](x) = \int_{-\beta_{l}}^{\beta_r} \dfrac{  f_{\mbox{\scriptsize cor}}(\mbox{\textbf{P}}_{\sharp,0})(\beta,u_{\beta}) R^2 (R\sin\beta+x)^k}{(R^2+u_{\beta}^2)^{\frac{1}{2}} (R\cos\beta - y_0)^{k+1}} d\beta}
\end{equation}
Here, as seen in Figure \ref{cons_trun}, $y_0$ is the height of a line $L$ lying inside the ROI but not intersecting with a scanned object, $\beta_l=\cos^{-1}(y_0/R)$, $\beta_r = 2\pi - \beta_l$, and $u_{\beta}= (x\cos\beta + y_0\sin\beta)/(R+x\sin\beta -y_0 \cos \beta)$.
Note that, for the artifact-free sinogram $\mbox{\textbf{P}}_*$ (i.e., consistent sinogram),  $\mathcal T_{k,y_0}[\mbox{\textbf{P}}_{*,0}](x)$ is a $k$-th order polynomial, where $\mbox{\textbf{P}}_{*,0}(\beta,u)=\mbox{\textbf{P}}_*(\beta,u,0)$ (see Figure \ref{cons_trun}). Hence, if $f_{\mbox{\scriptsize cor}}(\mbox{\textbf{P}}_{\sharp,0})=\mbox{\textbf{P}}_{*,0}$ (i.e., ideal sinogram correction), $\mathcal T_{k,y_0}[f_{\mbox{\scriptsize cor}}(\mbox{\textbf{P}}_{\sharp,0})]$ satisfies
\begin{equation}
\dfrac{\partial^{k+1}}{\partial x^{k+1}} \mathcal T_{k,y_0}[f_{\mbox{\scriptsize cor}}(\mbox{\textbf{P}}_{\sharp,0})](x) = 0, ~~ \forall ~ k = 0,1,2, \cdots
\end{equation}
This motivates the minimization problem \eqref{mini_para}.

In practice, this method can not be directly used and should be greatly simplified. To reduce the number of unknowns, the proposed method uses only zero order condition (i.e., $k_*=0$) and the following simplified approximation \cite{lee2019} with only four parameters $\boldsymbol \lambda = (\lambda_0,\lambda_1,\lambda_2,\lambda_3)$:
\begin{equation} \label{directcorrec}
f_{\mbox{\scriptsize cor}}(\mbox{\textbf{P}}_\sharp) = \left \lbrace \begin{array}{cl} \mbox{\textbf{P}}_\sharp  & \mbox{ if } \mbox{\textbf{P}}_\sharp \leq \Lambda \\
h_{\lambda_0,\Lambda}(\mbox{\textbf{P}}_\sharp)+  \displaystyle \sum_{i=1}^{3}\lambda_i (\mbox{\textbf{P}}_\sharp-\Lambda)^{i} & \mbox{ if } \mbox{\textbf{P}}_\sharp > \Lambda \end{array} \right.
\end{equation}
where $\Lambda$ is a sutiably chosen constant and
\begin{equation}\label{ht}
h_{\lambda_0,\Lambda}(t)=\dfrac{\lambda_0\Lambda-1}{2\lambda_0 e^{-\lambda_0\Lambda}}e^{-\lambda_0 t}+\dfrac{\lambda_0\Lambda+1}{2\lambda_0 e^{\lambda_0\Lambda}}e^{\lambda_0 t}
\end{equation}
That is, $f_{\mbox{\scriptsize cor}}$ in \eqref{directcorrec} is determined by
\begin{equation}
\boldsymbol \lambda = \underset{\boldsymbol \lambda}{\mbox{argmin}}  \int
\left| \dfrac{\partial}{\partial x} \mathcal T_{0,y_0}[f_{\mbox{\scriptsize cor}}(\mbox{\textbf{P}}_{\sharp,0})](x) \right|^2 dx
\end{equation}

\begin{figure*}[t]
	\centering
	\includegraphics[width=1\textwidth]{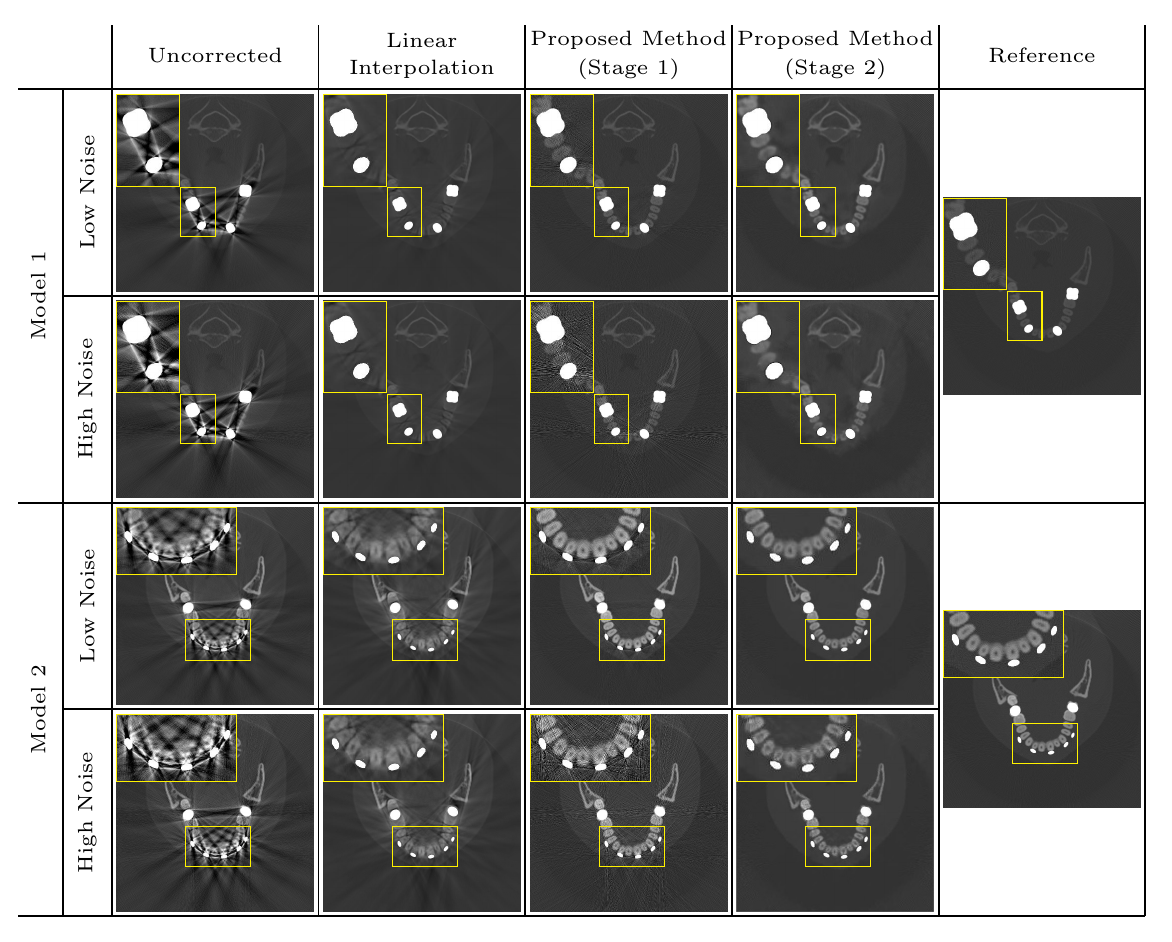}
	\caption{Numerical model experiment results. Model 1 simulates patient with metal implants and Model 2 resembles patient with metal implants along with brackets. From each model, sinogram data is generated with different noise levels. Each column displays reconstruction results by FDK algorithm (first), linear interpolation + FDK algorithm (second), and the proposed method (third and fifth), and a reference image (fifth). All images are displayed in the window is [C=500HU/W=5000HU].}	
	\label{numericalexp}
\end{figure*}

Owing to the sinogram inconsistency correction, we obtain the artifact-reduced image $\boldsymbol \mu = \mathcal R_{\diamond}^{{-1}} \circ f_{\mbox{\scriptsize cor}}(\mbox{\textbf{P}}_{\sharp})$. Here, the sinogram extrapolation method \cite{tisson2006} is additionally used to reduce cupping artifacts caused by FOV truncation.
Unfortunately, as seen in the fourth column of Figure \ref{numericalexp}, the sinogram correction tends to amplify noise-induced streaking artifacts, while reducing beam hardening-induced artifacts. Fortunately, the corrected image $\mathcal R_{\diamond}^{{-1}} \circ f_{\mbox{\scriptsize cor}}(\mbox{\textbf{P}}_{\sharp})$  has a more deep-learning friendly image than the uncorrected image $\mathcal R^{-1}_{\diamond} \mbox{\textbf{P}}_{\sharp}$, as shown in the second and fourth column of Figure \ref{numericalexp}.

\subsubsection{Deep learning process}
The deep learning process is designed to suppress noise-induced streaking artifacts in $\mathcal R_{\diamond}^{{-1}} \circ f_{\mbox{\scriptsize cor}}(\mbox{\textbf{P}}_{\sharp})$. The proposed method uses the convolutional neural network U-net \cite{ronneberger2015}, which is known to effectively reduce streaking artifacts \cite{jin2017,hyun2020,hyun2020-sol}.

Let $\{ (\boldsymbol \mu_\sharp^{(k)}, \boldsymbol \mu_*^{(k)}) \}_{k=1}^{N}$ be a training dataset, where $\boldsymbol \mu_\sharp^{(k)}$
is a noisy 3D CT image reconstructed by the sinogram inconsistency correction and $\mu_*^{(k)} = \mathcal R^{-1}_{\diamond}\mbox{\textbf{P}}_*^{(k)}$ is the corresponding noise-reduced image. The function $f_{\mbox{\scriptsize dl}}$ can be learned by the following training process:
\begin{equation}
f_{\mbox{\scriptsize dl}} = \underset{f_{\mbox{\scriptsize dl}} \in \mathbb{U}_{\mbox{\scriptsize net}}}{\mbox{argmin}} \sum_{k=1}^{N} \sum_{j=1}^{M} \| f_{\mbox{\scriptsize dl}}(\boldsymbol \mu_\sharp^{(k,j)}) - \boldsymbol \mu_*^{(k,j)}\|_{\ell^2}^2
\end{equation}
where $\mathbb{U}_{\mbox{\scriptsize net}}$ is a set of all learnable functions from U-net, $\boldsymbol \mu_\sharp^{(k,j)}$ is the $j$-th slice of $\boldsymbol \mu_\sharp^{(k)}$ on $z$-axis, $\boldsymbol \mu_*^{(k,j)}$ is the $j$-th slice of $\mathcal R_{\diamond}^{-1} \mbox{\textbf{P}}_*^{(k)}$ on $z$-axis, and $M$ is the total number of $z$-axis slices of $\boldsymbol \mu_\sharp$ and $\boldsymbol \mu_*$.

The overall structure of the U-net is described in Figure \ref{dl_process}. The architecture of the U-net comprises two parts; contracting and expansive path. Extracting feature maps from an input image, the contracting path is a repeated application of a $3\times3$ convolution with a rectified linear unit (ReLU) activation function and max-pooling. In the expansive path, a $3\times3$ convolution with ReLU and an average un-pooling is repeatedly applied and each un-pooled output is concatenated with the corresponding feature map in the contracting path to prevent loss of detailed information in the image. In the last layer of the expansive path, $1 \times 1$ convolution is applied to integrate the multi-channel feature map into one output.

\begin{table*}[t]
	\caption{\footnotesize{Quantitative comparison for numerical models }}
	\label{table1}
	\centering
	\begin{tabular}{C{0.75cm}|C{1.5cm} | C{2cm} C{2cm} C{2cm} C{2cm}}
		\hline\hline
		&\scriptsize{NRMSD($\%$)}&\scriptsize{Uncorrected}&\scriptsize{Linear Interpolation}&\scriptsize{Proposed Method (Stage 1)} &\scriptsize{Proposed Method (Stage 2)} \\
		\hline
		\multirow{2}{*}{\rotatebox{0}{\scriptsize{Model 1}}}&\scriptsize{Low Noise}&143.22&79.18&78.26&62.45\\
		\cline{2-6}
		&\scriptsize{High Noise}&223.24&81.57&251.14&78.80\\
		\hline\hline
		\multirow{2}{*}{\rotatebox{0}{\scriptsize{Model 2}}}&\scriptsize{Low Noise}&63.51 &58.25&46.59&40.98\\
		\cline{2-6}
		&\scriptsize{High Noise}&112.91 &59.47&151.18&55.42\\
		\hline\hline
	\end{tabular}
	\begin{tabular}{C{0.75cm}|C{1.5cm} | C{2cm} C{2cm} C{2cm} C{2cm}}
		\hline\hline
		&\scriptsize{SSIM}&\scriptsize{Uncorrected}&\scriptsize{Linear Interpolation}&\scriptsize{Proposed Method (Stage 1)} &\scriptsize{Proposed Method (Stage 2)} \\
		\hline
		\multirow{2}{*}{\rotatebox{0}{\scriptsize{Model 1}}}&\scriptsize{Low Noise}&0.93&0.97&0.97&0.98\\
		\cline{2-6}
		&\scriptsize{High Noise}&0.90&0.96&0.92&0.98\\
		\hline\hline
		\multirow{2}{*}{\rotatebox{0}{\scriptsize{Model 2}}}&\scriptsize{Low Noise}&0.94 &0.95&0.97&0.97\\
		\cline{2-6}
		&\scriptsize{High Noise}&0.90 &0.95&0.88&0.97\\
		\hline\hline
	\end{tabular}
\end{table*}

\begin{figure*}[t]
	\centering
	\includegraphics[width=1\textwidth]{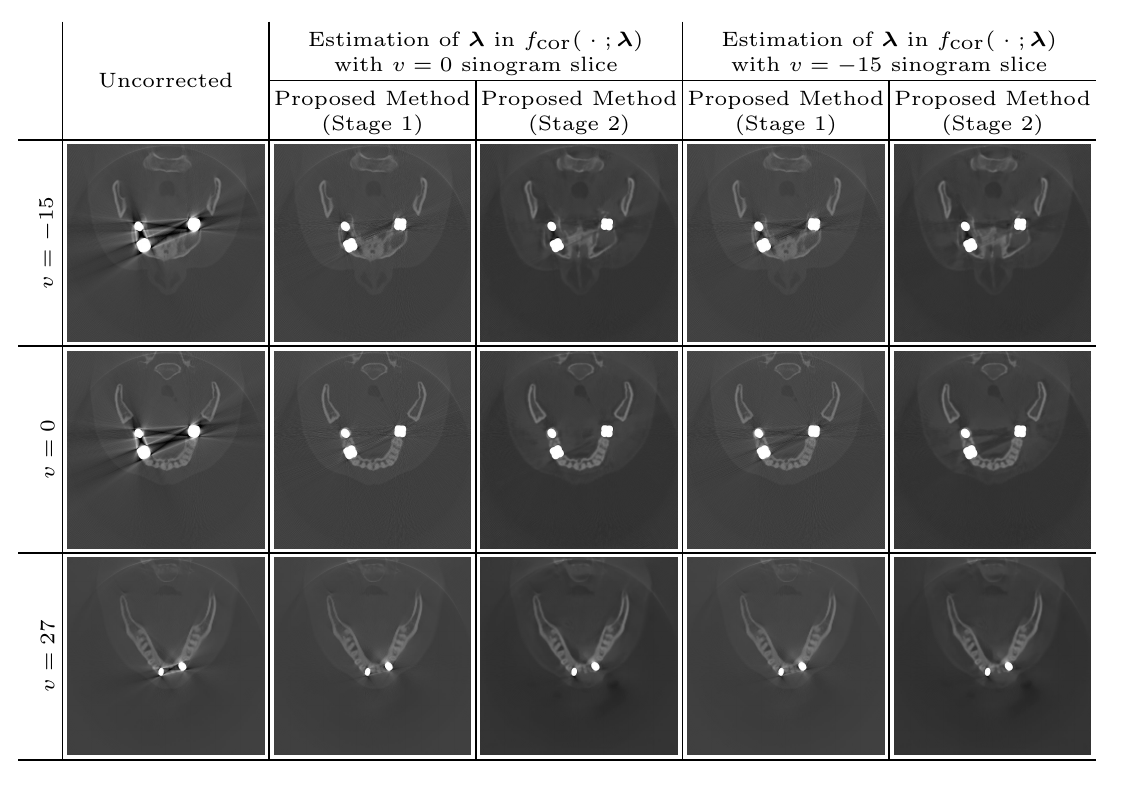}
	\caption{Several $z$-axis slice images of a 3D reconstruction image by the proposed method with a parameter estimation using a center ($v=0$) and a non-center ($v=-15$) sinogram slice. In the first column, three $z$-axis slices of an uncorrected image are displayed. The other columns display three $z$-axis slices of a reconstructed image by the proposed method using the parameter estimation \eqref{mini_para} with the center (second and third column) and the non-center slice (fourth and fifth column).	All images are displayed in the window [C=500HU/W=5000HU].}	
	\label{numericalexp2}
\end{figure*}

\begin{figure*}[t]
	\centering
	\includegraphics[width=1\textwidth]{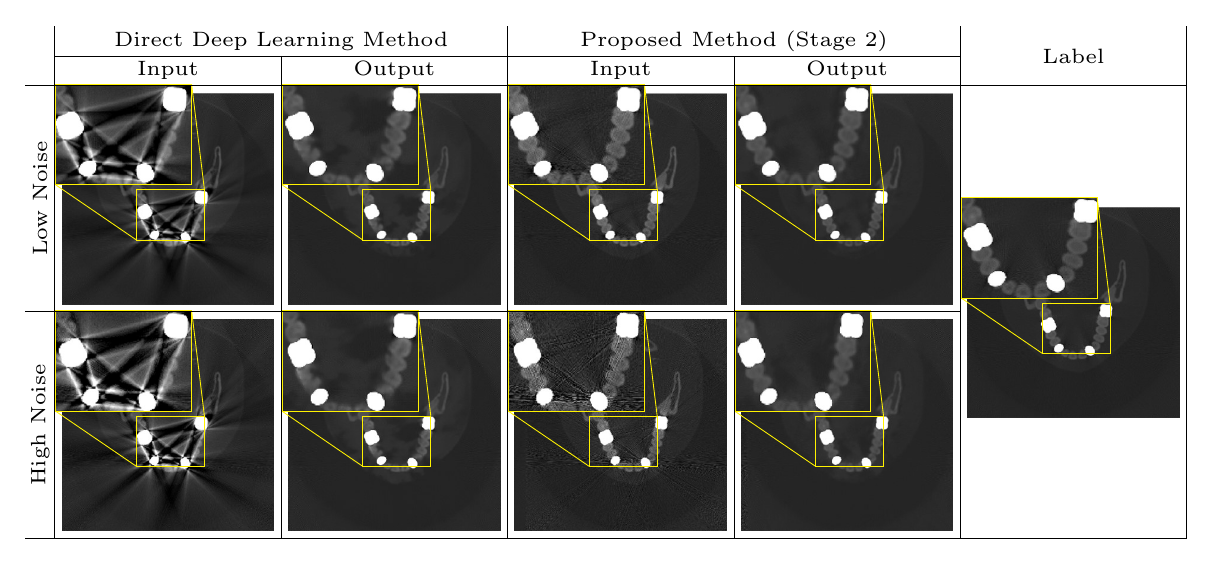}
	\caption{Comparison between the proposed method and a deep learning method that directly uses U-net to learn a function $f : \mathcal R^{-1}_{\diamond} \mbox{\textbf{P}} \mapsto \boldsymbol \mu_{*}$. The figure shows two experimental results for the case of low (first row) and high noise (second row). In each column, an input image (first and third), a deep learning output image (second and fourth), and a label image (fifth) are displayed.}	
	\label{DLcomparision}
\end{figure*}

\section{Result}
\subsection{Experimental Setting}
To evaluate the capability of the proposed method, several numerical simulations and real phantom experiments were performed in a MATLAB environment. Real phantom experiments were conducted using a commercial CBCT machine (Q-FACE, HDXWILL, South Korea). All numerical simulations were designed on the same scale and detector configuration according to those of a real CBCT machine. For cone-beam projection and back-projection, we modified the open-source algorithm \cite{kim2016}.

An acquired 3D sinogram comprises $658$ sinogram slices with respect to the $v$-axis whose size is $720 \times 654$. Here, $720$ is the number of projection views sampled uniformly in $[0,2\pi)$ and $654$ is the number of samples measured by the detector for each projection view. Among $654$ samples, $605$ samples were measured in the larger arm of the offset detector. In the reconstruction process, a sinogram was converted by the standard FDK algorithm into a CT image voxel of size $800 \times 800 \times 400$.

For deep learning, a training dataset was generated as follows. Metallic objects were inserted in metal-free images, by virtue of which simulated sinograms were obtained. In each metal-free image, many simulated sinograms can be generated by varying the shape and type of the inserted metal objects. After applying the sinogram inconsistency correction to each simulated sinogram, a set of training input data was obtained. All deep learning implementations were performed in a Pytorch environment \cite{paszke2019} with a computer system equipped with two Intel(R) Xeon(R) CPU E5-2630 v4, 128GB DDR4 RAM, and four NVIDIA GeForce GTX 2080ti GPUs. All training weights were initialized by a zero-centered normal distribution with a 0.01 standard deviation and a loss function was minimized using the Adam optimizer \cite{kingma2014}. Batch normalization was applied to achieve fast convergence and to mitigate the overfitting issue \cite{ioffe2015}.

\subsection{Numerical Simulation}
In the numerical simulation, a sinogram was generated by inserting metal materials in the metal-free CT human head image voxel and by adding Poisson and electric noise. We referred to the attenuation coefficient values of metal implants in \cite{hubbell1995} and polychromatic X-ray energy spectrums in \cite{mahesh2013}.

To test the proposed method, two numerical models (Model 1 and Model 2) were designed. Each model is generated by placing metallic objects resembling a dental implant (Model 1) and bracket (Model 2). In each model, two sinograms (low noise and high noise) were generated with two different ampere settings.

Figure \ref{numericalexp} shows results of beam hardening artifact reduction by using linear interpolation, and the proposed method. Quantitative comparisons of these methods, based on normalized root mean square difference (NRMSD) and structural similarity (SSIM) \cite{wang2004} metrics, are listed in Table \ref{table1}. The linear interpolation method reduces beam hardening artifacts, whereas it destroys the morphological structure of tooth. In contrast, the proposed method reduces not only beam hardening artifacts, but also improves the quality of the tooth image.

In high noise case, the advantage of using U-net is emphasized. In stage 1 of the proposed method, tooth structure are considerably improved, whereas noise-related streaking artifacts are amplified (see Figure \ref{numericalexp}). Owing to the amplified artifacts, the proposed method achieves poor quantitative evaluation results in the stage 1, as seen in Table \ref{table1}. In the stage 2, however, as shown in Figure \ref{numericalexp} and Table \ref{table1}, U-net successfully suppresses the streaking artifacts and, therefore, the proposed method shows good quantitative performance.

Several $z$-axis slices of a 3D reconstruction image are provided in Figure \ref{numericalexp2}. The proposed method alleviates beam hardening artifacts in the entire image domain. We also compare the reconstruction performance of the proposed method when using a different sinogram slice in the parameter estimation \eqref{mini_para}. Even with the sinogram slice at $v=-15$, the proposed method fairly reduces beam hardening artifacts; however its performance is worse than that obtained previously.

In Figure \ref{DLcomparision}, we compare the proposed method with the deep learning method that directly uses an uncorrected image (i.e. $\mathcal R_{\diamond}^{-1} \mbox{\textbf{P}}$) as an input of U-net. Compared to the direct application of U-net, the proposed method has the advantage of recovering the tooth structure. This is because the sinogram inconsistency correction makes tooth feature in a deep learning input image salient.

\subsection{Phantom experiments}
\begin{figure*}[t]
	\centering
	\includegraphics[width=1\textwidth]{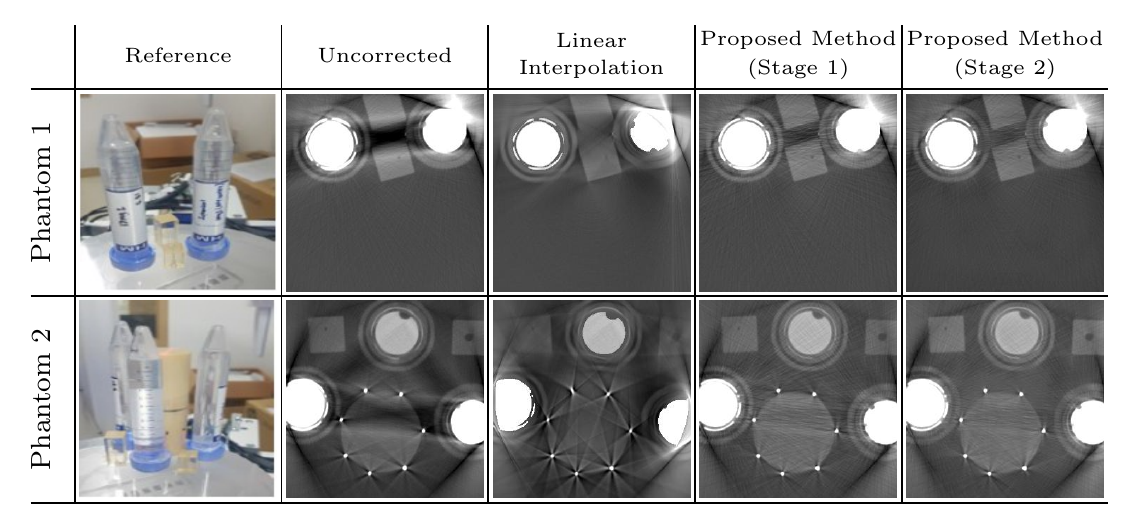}
	\caption{Real phantom experimental results. Phantom 1 includes acryl blocks and 2 cylinders filled with a fluid iodine concentration of 370 mgI/ml and Phantom 2 contains acryl block, tissue-equivalent phantom, root canal filling and cylinders filled with different fluid. Each column displays a reference image (first) or a reconstruction results by FDK algorithm (second), linear interpolation + FDK algorithm (third), and the proposed method (fourth and fifth). All images are displayed in the window is [C=0HU/W=4000HU].}
	\label{realexp}
\end{figure*}
\begin{figure*}[t]
	\centering
	\includegraphics[width=0.7\textwidth]{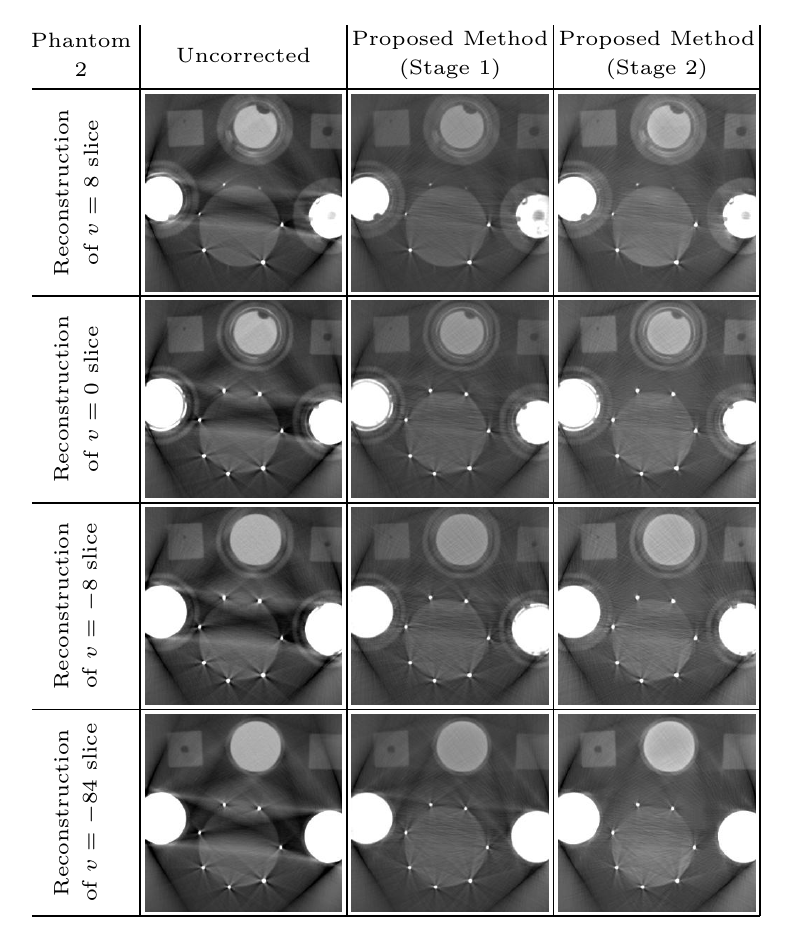}
	\caption{Several $z$-axis slice images of a 3D reconstruction image by the proposed method (second and third row). All images are displayed in the window is [C=0HU/W=4000HU]}	
	\label{otherslice}
\end{figure*}
For real phantom experiments, a real dental CBCT machine (Q-FACE, HDXWILL, Seoul, South Korea) was used with a tube voltage 90kVp, tube current of 10mA, and Cu filtration of 0.5mm. Two phantom models were constructed using acryl block, tissue-equivalent phantom, root canal filling, and cylinders filled with a high attenuating fluid (iodinated contrast media; Dongkook Pharma, Seoul, South Korea). Figures \ref{realexp} and \ref{otherslice} present the reconstruction results for the real phantom experiments. It is observed that the proposed method significantly reduces beam hardening artifacts while improving the image quality of scanned objects in the entire image domain.

\section{Discussion and Conclusion}
\begin{figure*}[t]
	\centering
	\includegraphics[width=1\textwidth]{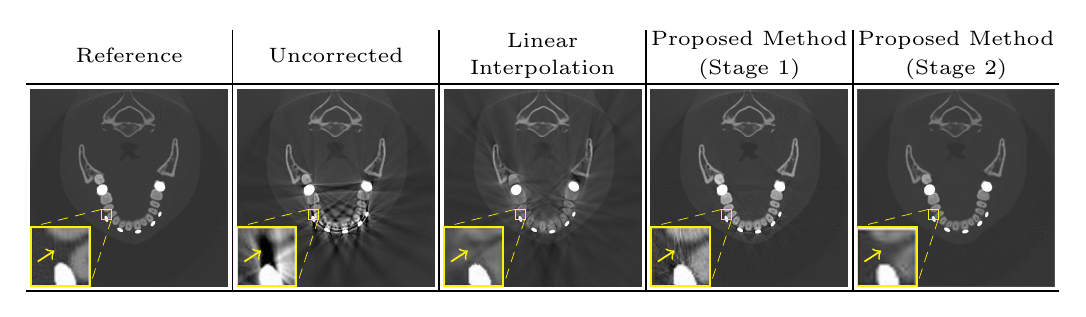}
	\caption{Beam hardening artifact reduction results using a numerical model. Each column display a reconstructed image by FDK algorithm (first), linear interpolation + FDK algorithm (second), and the proposed method (third and fourth), and a reference (fifth). As shown in yellow boxes of Figure, the proposed method in stage 1 alleviates beam hardening artifacts related with metal-teeth interaction, but it amplifies noise-related streaking artifacts. After taking DL, the streaking artifacts are reduced.}		
	\label{interactionrduce}
\end{figure*}

This paper proposes a beam hardening reduction method for low-dose dental CBCT that overcomes the hurdles caused by the offset detector configuration and the interior-ROI-oriented scan. The proposed BHC method is a two-step method. In the first step, the sinogram corrector $f_{\mbox{\scriptsize cor}}$ in Section \ref{secSIC} was applied to reveal the tooth structure that is obscured by the beam hardening artifacts because of the sinogram inconsistency. Unfortunately, this sinogram corrector tends to amplify noise-related streaking artifacts. To curb this, at the second stage, these noise-related artifacts are significantly eliminated through the deep learning method.

Numerical and real phantom experiments were performed to show that the proposed method is successfully applied in dental CBCT environment. It is further observed that the proposed method can effectively deal with beam hardening artifacts related to not only metallic objects but also metal-bone and metal-teeth interaction, as shown in Figure \ref{interactionrduce}.

\EOD
\end{document}